\documentclass[aps,prb,reprint,superscriptaddress,nofootinbib]{revtex4-2}

\usepackage{graphicx}
\usepackage{dcolumn}
\usepackage{bm}
\usepackage[version=4]{mhchem}
\usepackage{booktabs}
\usepackage{comment}
\usepackage{braket}
\usepackage{color}
\usepackage{advdate}

\usepackage[dvipdfmx]{hyperref}
\hypersetup{
 colorlinks=true,
 linkcolor=blue,
 citecolor=blue,
 urlcolor=blue,
}

\begin{document}

\title{Observation of field-induced single-ion magnetic anisotropy in a multiorbital Kondo alloy (Lu,Yb)Rh$_2$Zn$_{20}$}

\author{T. Kitazawa}
\email[takafumi.kitazawa.s5@dc.tohoku.ac.jp]{}
\affiliation{Graduate School of Science, Tohoku University, Sendai 980-8578, Japan}
\affiliation{Advanced Science Research Center, Japan Atomic Energy Agency, Tokai, Ibaraki 319-1195, Japan}

\author{Y. Ikeda}
\affiliation{Institute for Materials Research, Tohoku University, Sendai 980-8577, Japan}
\author{T. Sakakibara}
\affiliation{The Institute for Solid State Physics, The University of Tokyo, Kashiwa, Chiba 277-8581, Japan}
\author{A. Matsuo}
\affiliation{The Institute for Solid State Physics, The University of Tokyo, Kashiwa, Chiba 277-8581, Japan}
\author{Y. Shimizu}
\affiliation{Institute for Materials Research, Tohoku University, Oarai, Ibaraki 311-1313, Japan}
\author{Y. Tokunaga}
\affiliation{Advanced Science Research Center, Japan Atomic Energy Agency, Tokai, Ibaraki 319-1195, Japan}
\author{\\Y. Haga}
\affiliation{Advanced Science Research Center, Japan Atomic Energy Agency, Tokai, Ibaraki 319-1195, Japan}
\author{K. Kindo}
\affiliation{The Institute for Solid State Physics, The University of Tokyo, Kashiwa, Chiba 277-8581, Japan}
\author{Y. Nambu}
\affiliation{Institute for Materials Research, Tohoku University, Sendai 980-8577, Japan}
\affiliation{Fusion Oriented Research for Disruptive Science and Technology, Japan Science and Technology Agency, Kawaguchi, Saitama 332-0012, Japan}
\affiliation{Organization for Advanced Studies, Tohoku University, Sendai 980-8577, Japan}
\author{K. Ikeuchi}
\affiliation{Neutron Science and Technology Center, Comprehensive Research Organization for Science and Society (CROSS), Tokai, Ibaraki 319-1106, Japan}
\author{K. Kamazawa}
\affiliation{Neutron Science and Technology Center, Comprehensive Research Organization for Science and Society (CROSS), Tokai, Ibaraki 319-1106, Japan}
\author{M. Ohkawara}
\affiliation{Institute for Materials Research, Tohoku University, Sendai 980-8577, Japan}
\author{M. Fujita}
\affiliation{Institute for Materials Research, Tohoku University, Sendai 980-8577, Japan}

\date{\AdvanceDate[-1]\today}

\begin{abstract}
We demonstrate field-induced single-ion magnetic anisotropy resulting from the multiorbital Kondo effect on the diluted ytterbium alloy (Lu$_{1-x}$Yb$_x$)Rh$_2$Zn$_{20}$. Single-ion anisotropic metamagnetic behavior is revealed in low-temperature regions where the local Fermi-liquid state is formed. Specific heat, low-field magnetic susceptibility, and resistivity indicate reproduction of the ground-state properties by the SU($N=8$) Kondo model with a relatively large $c$-$f$ hybridization of $T_\mathrm{K} = 60.9 \ \mathrm{K}$.  Dynamical susceptibility measurements on YbRh$_2$Zn$_{20}$ support realizing the multiorbital Kondo ground state in (Lu$_{1-x}$Yb$_x$)Rh$_2$Zn$_{20}$. 
The single-ion magnetic anisotropy becomes evident above $\sim$$5 \ \mathrm{T}$, which is lower than the isotropic Kondo crossover field of 22.7 T, verifying blurred low-lying crystal field states through the multiorbital Kondo effect.
\end{abstract}

\maketitle

\section{Introduction}

The Kondo effect \cite{Kondo1964, Hewson1993}, which is a verifiable example of asymptotic freedom in many-body electron systems, has provided deep insights into various fields of condensed matter \cite{Coleman2007} and high-energy physics \cite{Hattori2015}.
A prototypical case of the Kondo problem for $S=1/2$ and the generalized SU($N$) model \cite{Anderson1961, Coqblin1969}, where a local magnetic impurity in the Fermi sea has total degeneracy $N$ for the spin and orbital degrees of freedom, has been studied intensively and solved exactly in the 1980s (for example, see Refs. \cite{Andrei1980, Schlottmann1983}).
In the last 40 years, many studies have investigated nontrivial Kondo problems such as overscreening and underscreening cases for the multichannel model with complex internal degrees of freedom \cite{Nozieres1980}. This includes investigating a non-Fermi-liquid fixed point in a two-channel (quadrupole) Kondo model \cite{Cox1987, Yamane2018} and the Kondo effect on multiorbital (multilevel) systems in quantum dots and organic molecules adsorbed onto a metal surface \cite{Jarillo-Herrero2005, Keller2014, Ferrier2017, Minamitani2012}.

This study investigates multiorbital Kondo effects in bulk metallic systems, 
focusing on the distinctive physical properties. 
In a multiorbital system, the $c$-$f$ hybridization effect may activate a nontrivial degree of freedom, such as a higher-rank multipole, and unconventional quantum phenomena with various physical properties as discussed in heavy-fermion physics \cite{Martelli2019, Amorese2020, Shu2021, Khim2021, Hafner2022}. 
In this paper, we demonstrate a prominent feature, i.e., field-induced single-ion magnetic anisotropy, originating from the multiorbital Kondo effect on an Yb-based compound (Lu$_{1-x}$Yb$_x$)Rh$_2$Zn$_{20}$. 
The enhancement of the magnetic anisotropy is probably caused by the deformation of the ground-state wave function from an almost isotropic multiorbital Kondo singlet state to an anisotropic electronic state.

We first demonstrate that the ground-state properties of (Lu$_{1-x}$Yb$_{x}$)Rh$_2$Zn$_{20}$ ($x=0.014$) can be recognized as a multiorbital Kondo ground state and qualitatively reproduced by the SU($N=8$) Kondo model with only one parameter. 
Next, we demonstrate and discuss the evolution of single-ion magnetic anisotropy at high magnetic fields. 
We examine purely single-site Kondo behavior to corroborate that our findings come from a single-site effect rather than an intersite phenomenon. 
Hence, the influence of intersite magnetic interactions between Yb ions was excluded by diluting magnetic Yb ions by $\sim$1\% with nonmagnetic Lu ions.  
Hereinafter, we denote (Lu$_{0.986}$Yb$_{0.014}$)Rh$_2$Zn$_{20}$ as $x=0.014$ for simplicity.

\section{Experimental details}

Single crystals of (Lu$_{1-x}$Yb$_x$)Rh$_2$Zn$_{20}$ were grown using the Zn self-flux method with the nominal composition of $(\mathrm{Lu}$ $+$ $\mathrm{Yb}) : \mathrm{Rh} : \mathrm{Zn} = 1 : 2 : 97$ for $X = 0$, 0.1, and 0.25 or $1 : 2 : 60$ for $X = 1$, where $X$ is the nominal Yb concentration \cite{Jia2008, Honda2013}. The constituent elements were melted at $1100 \ ^\circ \mathrm{C}$ and then cooled to $600-700 \ ^\circ \mathrm{C}$ at $-2 \ ^\circ \mathrm{C}/\mathrm{h}$ for $X = 0$, $0.1$, and $1$ or $-1.5 \ ^\circ \mathrm{C}/\mathrm{h}$ for $X = 0.25$. 
The actual Yb compositions of $X = 0$ and 0.1 samples were examined via the inductively coupled plasma atomic emission spectroscopy (ICP-AES) analysis, conducted under the support of the Analytical Research Core for Advanced Materials, Institute for Materials Research, Tohoku University. 
For $X = 0.25$, the actual Yb composition was estimated on the basis of the $M/H(T)$ data. 
The obtained Yb concentrations in $X = 0.1$ and 0.25 samples were $x = 0.014$ and 0.075, respectively.
In contrast, the value of $x$ in $X = 0$ samples was lower than $4.9 \times 10^{-4}$, which is the limit of quantitative analysis for ICP-AES. 
Therefore, $X = 0$ samples were used as a nonmagnetic reference compound for (Lu$_{1-x}$Yb$_x$)Rh$_2$Zn$_{20}$.
The details of the ICP-AES analysis for $X = 0$ and $0.1$ and $M/H(T)$ data for $X = 0.25$ are provided in Secs. SI and SIII C, respectively, of the Supplemental Material \cite{SM}.

The specific heat was measured using a physical property measurement system with a $^3$He cooling option (PPMS; Quantum Design).
The magnetization data were collected using a commercial dc superconducting quantum interference device (SQUID) magnetometer (MPMS; Quantum Design) and a high-resolution capacitive Faraday magnetometer  \cite{Sakakibara1994, Shimizu2021}.
The electrical resistivity was measured using the four-probe method with the PPMS.
The inelastic neutron scattering (INS) experiments in \ce{YbRh2Zn20} and \ce{LuRh2Zn20} were conducted using the Fermi chopper spectrometer 4SEASONS \cite{Kajimoto2011} at the Materials and Life Science Experimental Facility (MLF) of the Japan Proton Accelerator Research Complex (J-PARC).
For the INS experiments, powdered samples were prepared by crushing single-crystal samples of \ce{YbRh2Zn20} and \ce{LuRh2Zn20}, weighing 4.1 and 3.9 g, respectively, and placing each sample in a home-built Al double-cylinder cell.
The colossal event data were extracted using the Utsusemi software developed at MLF \cite{Inamura2013}.

\section{Results}

\begin{figure}[t]
	\includegraphics[keepaspectratio, width=8.6cm, clip]{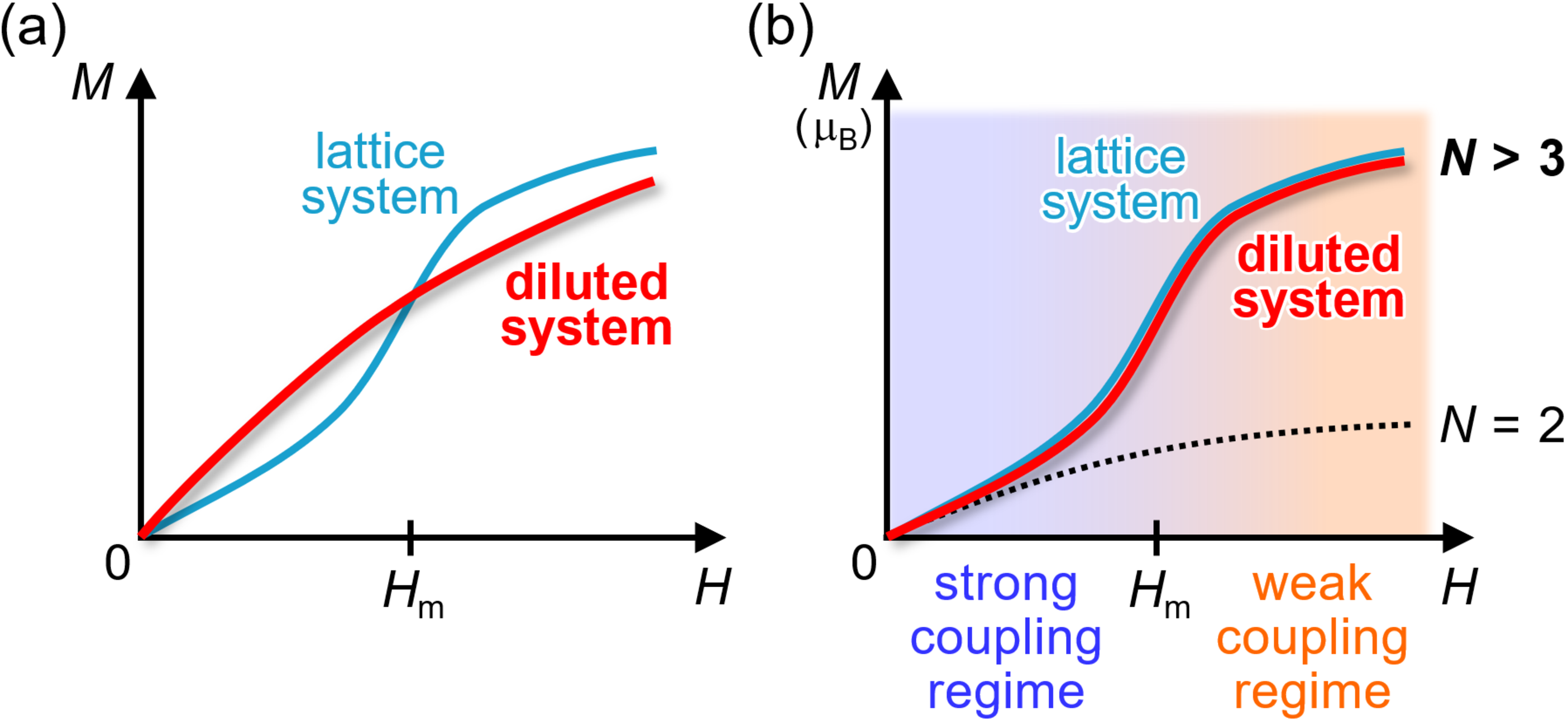}
	\caption{Schematic of metamagnetism originating from (a) intersite exchange interactions and (b) the multiorbital Kondo effect ($N>3$). The dotted line shows a typical SU(2) Kondo effect behavior. \label{magnetization_illustration}}
\end{figure}

As predicted by Hewson and other researchers \cite{Hewson1983_1, Hewson1983_2, Newns1987, Cox1987_2, Kang1996, Bickers1985, Cox1985, Bickers1987}, the essential ingredients of the SU($N>3$) Kondo effect are 
(i) local Fermi-liquid behavior, 
(ii) a nonlinear increase in magnetization in a dilute-limit system [single-site metamagnetic behavior, Fig. \ref{magnetization_illustration}(b)] \cite{metamag},  
(iii) a local maximum in the temperature-dependent static magnetic susceptibility, and 
(iv) a local minimum in the temperature-dependent reciprocal lifetime of the Kondo singlet state, evaluated from the half width at half maximum $\Gamma (T)$ of the energy-dependent dynamical magnetic susceptibility. 
In Secs. \ref{subsec_Fermi}-\ref{subsec_quasielastic}, we present experimental evidence of these ingredients sequentially.
In Sec. \ref{subsec_magnetic_anisotropy}, we introduce an unexpected phenomenon in the SU($N$) Kondo model: field-induced single-ion magnetic anisotropy.

\subsection{Local Fermi liquid}
\label{subsec_Fermi}

\begin{figure}[t]
	\includegraphics[keepaspectratio, width=8.6cm, clip]{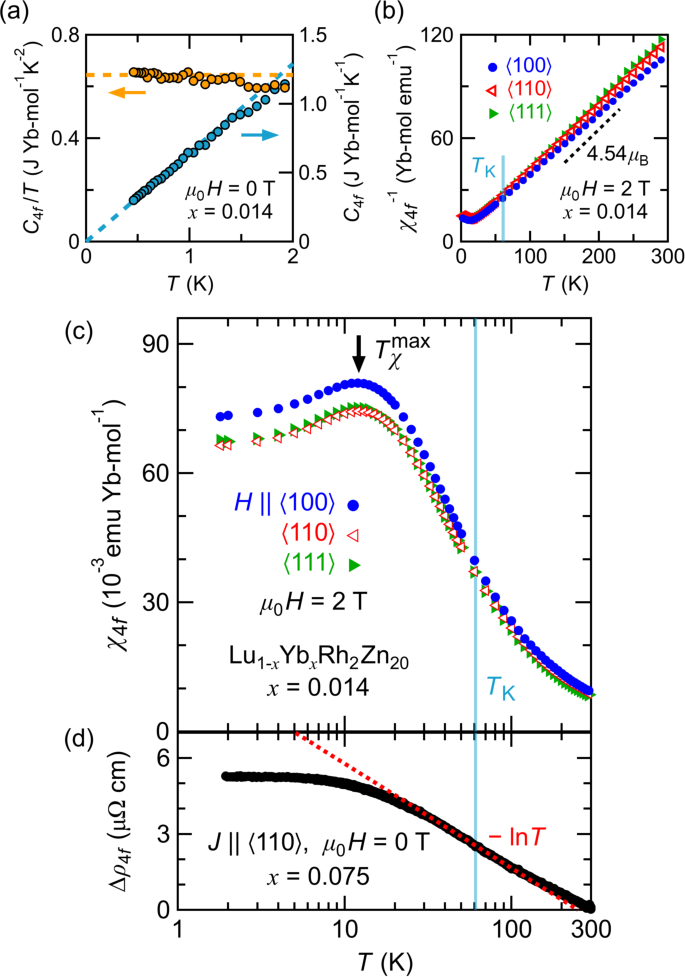}
	\caption{Temperature dependence of (a) specific heat $C_{4f}$ (right axis) and $C_{4f}T^{-1}$ (left axis), (b) reciprocal magnetic susceptibility ${\chi_{4f}}^{-1}$, and (c) magnetic susceptibility $\chi_{4f}$ of \ce{Lu_{0.986}Yb_{0.014}Rh2Zn20}. The dashed lines in (a) represent the fitting results with $C_{4f} = \gamma_{4f} T$ below $1 \ \mathrm{K}$. The light blue lines in (b), (c), and (d) represent $T_\mathrm{K} = 60.9 \ \mathrm{K}$. In (b), the dashed line shows the ideal slope for Yb$^{3+}$ with an effective magnetic moment of 4.54 $\mu_{\mathrm{B}}$. (d) Temperature dependence of the $4f$ contribution of resistivity with respect to that at $T = $ 300 K, $\Delta \rho_{4f}(T) = \rho_{4f}(T) - \rho_{4f}(T = 300 \ \mathrm{K})$, in Lu$_{0.925}$Yb$_{0.075}$Rh$_2$Zn$_{20}$. \label{LYRZ_macro_Tdep}
	}
\end{figure}

Here, we show evidence of local Fermi-liquid behavior in a single-crystalline $x=0.014$ sample. Figure  \ref{LYRZ_macro_Tdep}(a) displays the temperature dependence of the specific heat $C_{4f}$ (right axis) and $C_{4f}T^{-1}$ (left axis) of the $x=0.014$ sample. The 4$f$-electron contribution was extracted by subtracting the lattice specific heat ($= \beta T^3$) as a background \cite{SM}. The specific heat is proportional to $T$ at low temperatures, indicating a Fermi-liquid ground state. The electronic specific heat coefficient $\gamma_{4f}$ is evaluated as 645(2) mJ K$^{-2}$ mol$^{-1}$ per Yb ion.
The relatively large $\gamma_{4f}$ implies a large degree of freedom of the 4$f$ electron.

The local Fermi-liquid behavior is also observed in the static magnetic susceptibility in Figs. \ref{LYRZ_macro_Tdep}(b) and \ref{LYRZ_macro_Tdep}(c). 
The 4$f$-electron contribution toward the static magnetic susceptibility $\chi_{4f}$ asymptotically approaches a constant at low temperatures, where the localized magnetic moments of Yb ions are screened by conduction electrons as indicated by the specific heat data. 
$\chi_{4f}(T)$ was evaluated by subtracting slightly temperature-dependent paramagnetic and diamagnetic contributions in the solvent compound LuRh$_2$Zn$_{20}$ \cite{SM}. The evaluation reliability is supported by the $\chi_{4f}$ results at high temperatures. The reciprocal susceptibility ${\chi_{4f}}^{-1}$ shows Curie-Weiss behavior. Moreover, the evaluated effective magnetic moment $\mu_{\textrm{eff}}$, listed in Table \ref{tab:chi_property}, agrees reasonably well with 4.54 $\mu_{\textrm{B}}$ of Yb$^{3+}$ within $\sim$5\% accuracy in all directions.

Evident local Fermi-liquid behavior is also observed in the electrical resistivity in Fig. \ref{LYRZ_macro_Tdep}(d).
The $4f$ contribution of the resistivity with respect to that at $T = $ 300 K, $\Delta \rho_{4f}(T) = \rho_{4f}(T) - \rho_{4f}(T = 300 \ \mathrm{K})$, in Lu$_{0.925}$Yb$_{0.075}$Rh$_2$Zn$_{20}$, shows a logarithmic increase with decreasing temperature and saturation below $\sim$5 K, which exhibits the impurity Kondo singlet ground state. 
The interaction between magnetic impurities is negligible because of a lack of a local maximum in $\rho_{4f}(T)$ \cite{Larsen1978_1, Larsen1978_2}.
Besides, the observed local Fermi-liquid behavior in $C_{4f}(T)$, $\chi_{4f}(T)$, and $\rho_{4f}(T)$ excludes the influence of randomness, such as the distribution of Kondo temperature, for which non-Fermi-liquid behavior should be observed \cite{Dobrosavljevic1992, Andrade1998, Blanckenhagen2001, Kim2002}.
Therefore Yb impurities in $x = 0.014$ samples are considered isolated impurities.

\begin{table}[t]
\caption{\label{tab:chi_property}
Effective magnetic moment $\mu_\mathrm{eff}$, Curie-Weiss temperature $\theta_\mathrm{p}$, magnetic susceptibility $\chi_{4f}$ at $1.8 \ \mathrm{K}$, Wilson ratio $R_\mathrm{W}$, and $T_{\chi}^{ \mathrm{max}}$ for $H \ || \ \langle 100 \rangle$, $\langle 110 \rangle$, and $\langle 111 \rangle$ in \ce{Lu_{0.986}Yb_{0.014}Rh2Zn20}.}
\catcode`!=\active \def!{\phantom{.}}
\catcode`?=\active \def?{\phantom{0}}
\begin{ruledtabular}
\begin{tabular}{cccccc}
&
\begin{tabular}{c} $\mu_\mathrm{eff}$ \\ $(\mu_\mathrm{B}/\mathrm{Yb})$ \end{tabular}&
\begin{tabular}{c} $\theta_\mathrm{p}$ \\ $(\mathrm{K})$ \end{tabular}&
$\chi_{4f}(1.8 \ \mathrm{K})$\footnotemark[1]&
$R_\mathrm{W}$&
\begin{tabular}{c} $T_{\chi}^{\mathrm{max}}$ \\ $(\mathrm{K})$ \end{tabular}\\
\midrule
$\langle100\rangle$ & 4.763(5) & $-10.8(2)!$ & 73.1(1) & 1.093(5) & 12 \\
$\langle110\rangle$ & 4.61(3)? & $-10.2(9)!$ & 66.5(2) & 1.06(1)? & 12 \\
$\langle111\rangle$ & 4.51(4)? & $?-8.2(1.5)$ & 67.6(6) & 1.13(2)? & 12 \\
\end{tabular}
\end{ruledtabular}
\footnotetext[1]{The units are $\mathrm{10^{-3}emu \ (Yb \ mol)^{-1}}$.}
\end{table}

Orbital degeneracy of the local Fermi-liquid ground state, $N$, can be determined by examining the Wilson ratio $R_\mathrm{W} = \pi^2 k_{\mathrm{B}}^2 \chi_{4f}(0) / (\mu_{\textrm{eff}}^2 \gamma_{4f})$. Here, $k_{\mathrm{B}}$ is the Boltzmann constant, and $\chi_{4f}(0)$ is the magnetic susceptibility in the Fermi-liquid state at $T = 0$ K. We used the value of $\chi_{4f}$($T$) at 1.8 K as an approximate value of $\chi_{4f}(0)$. 
The evaluated values of $R_\mathrm{W}$, listed in Table \ref{tab:chi_property}, agree with the theoretically expected Wilson ratio $R_\mathrm{W} = N/(N-1) =(2J+1)/2J = 1.14$ for $N=8 \ (J=7/2)$ rather than 2 ($N = 2$), 1.33 ($N = 4$), and 1.2 ($N = 6$) \cite{Hewson1993, Hewson1983_2, WR}. 
The large orbital degeneracy of $N = 8$ can be verified using the results of ${\chi_{4f}}^{-1}(T)$ and $\Delta \rho_{4f}(T)$. 
First,  ${\chi_{4f}}^{-1}(T)$ displayed in Fig. \ref{LYRZ_macro_Tdep}(b) shows the Curie-Weiss behavior above $\sim$80 K (see also Fig. S5 in the Supplemental Material \cite{SM}), and the evaluated $\mu_{\textrm{eff}}$ listed in Table \ref{tab:chi_property} is close to the localized case of $4.54 \ \mu_{\textrm{B}}$ for Yb$^{3+}$, suggesting that $f$ electrons occupy all crystalline-electric field (CEF) levels with an equal probability above $\sim$80 K. 
Thus the overall CEF energy scale divided by $k_\mathrm{B}$, $\Delta_{\mathrm{CEF}}/k_\mathrm{B}$, is at most $\sim$80 K. 
Next, if $T_{\mathrm{K}} < \Delta_{\mathrm{CEF}}/k_{\mathrm{B}} \ (\lesssim 80 \ \mathrm{K})$, that is, $N \le 6$, $\rho_{4f}(T)$ below room temperature should exhibit a local maximum or a shoulderlike structure, reflecting the depopulation of $f$ electrons from the excited to lower CEF levels upon cooling \cite{Zlatic2005, Ocko2001, Nakatsuji2002, Kohler2008, Pikul2012, Lee2019}. However, $\Delta \rho_{4f}(T)$ in Fig. \ref{LYRZ_macro_Tdep}(d) shows no such anomaly; thus $k_{\mathrm{B}}T_{\mathrm{K}} > \Delta_{\mathrm{CEF}}$ and $N = 8$.

$T_{\mathrm{K}}$ of the $x=0.014$ sample is evaluated as 60.9 K with a typical relation between $T_{\mathrm{K}}$ and $\gamma_{4f}$ for $N=8$ \cite{Rajan1983, Kaihe2005, TK}.
Although the detailed CEF energy structure is uncertain, an upper boundary of the CEF splitting can be estimated to be $\sim$60 K because of $k_{\mathrm{B}}T_{\mathrm{K}} > \Delta_{\mathrm{CEF}}$ and $T_{\mathrm{K}} = 60.9 \ \mathrm{K}$. For a small CEF energy scale in the $LT_{2}$Zn$_{20}$ family ({\it L}, lanthanoids; {\it T}, transition metals), a topologically close-packed phase, also known as the Frank-Kasper cage structure, is crucial in suppressing the CEF level splitting owing to an almost spherical coordinate of the Zn atoms around the {\it L} ion \cite{Torikachvili2007, Onimaru2016}. The CEF energy scale $\Delta_{\mathrm{CEF}}/k_{\mathrm{B}}$ was evaluated as $\sim$70 K at most for the parent compound YbRh$_2$Zn$_{20}$ \cite{Honda2013} and $\sim$30 K for the isovalent compound YbCo$_2$Zn$_{20}$ \cite{Takeuchi2011, Kaneko2012}.  The evaluated $\Delta_{\mathrm{CEF}}$ for YbRh$_2$Zn$_{20}$ is supported by our neutron-scattering experiments, where no well-defined CEF excitation was observed, as mentioned in Sec. \ref{subsec_quasielastic}. 

\subsection{Single-site metamagnetic behavior}
\label{subsec_metamag}

\begin{figure}[t]
	\includegraphics[keepaspectratio, width=8.6cm, clip]{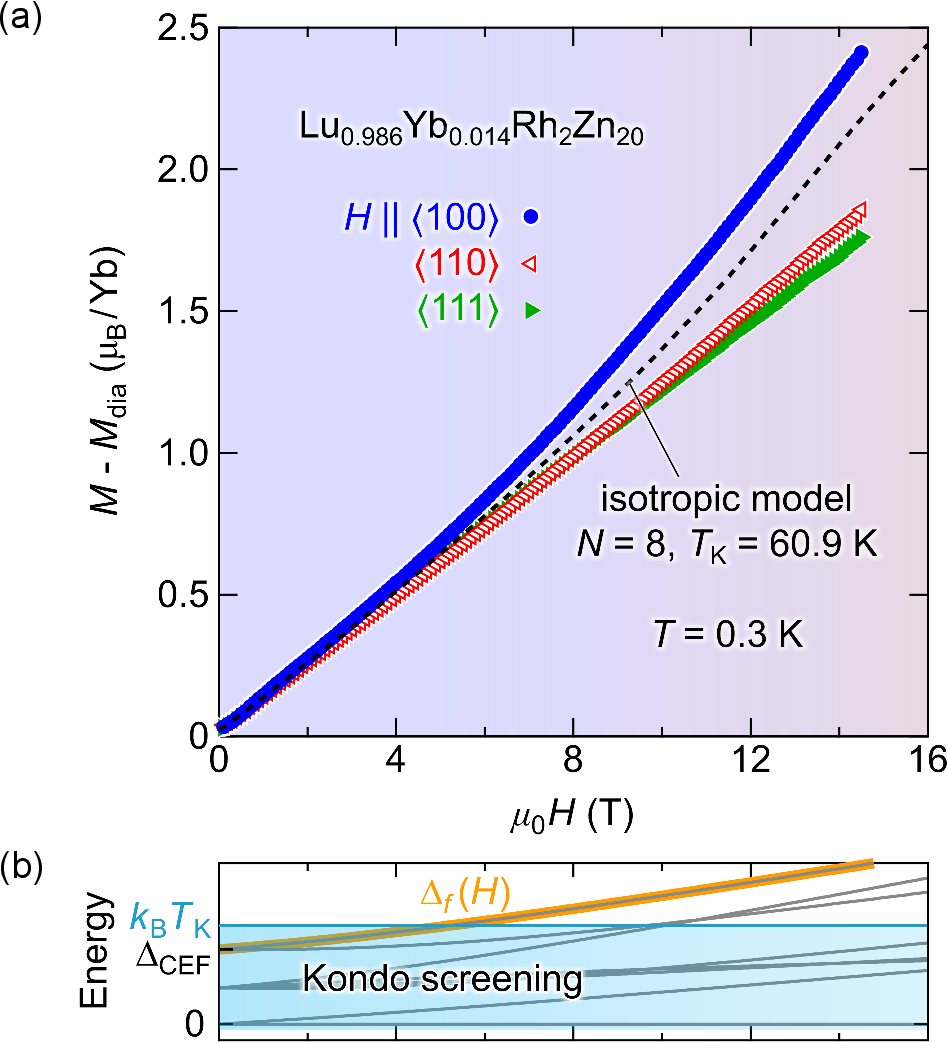}
	\caption{ 
	(a) Magnetic field dependence of magnetization on \ce{Lu_{0.986}Yb_{0.014}Rh2Zn20} measured at 0.3 K for $H \ || \ \langle 100 \rangle$, $\langle 110 \rangle$, and $\langle 111 \rangle$. The diamagnetic contribution $M_\mathrm{dia}$ of LuRh$_2$Zn$_{20}$ was subtracted. The dashed line in (a) is the theoretical curve based on the Coqblin-Schrieffer model \cite{Hewson1983_2} for $N = 8$ and $T_\mathrm{K} = 60.9 \ \mathrm{K}$ ($T_1 = 17.7 \ \mathrm{T}$). (b) Schematic of the magnetic field evolution of the eight $f$ levels (gray lines) and characteristic energy scale $\Delta_f (H)$ (yellow line), defined as the difference between the highest and lowest energy in the CEF levels. $\Delta_f (H = 0)$ corresponds to the overall CEF energy scale $\Delta_\mathrm{CEF}$. The energy width of the blue shaded region represents $k_\mathrm{B}T_\mathrm{K}$.
	\label{LYRZ_Magnetization}
	}
\end{figure}

Next, we show the second ingredient of the SU($N$) Kondo ground state: single-site metamagnetic behavior. As seen in Fig. \ref{LYRZ_Magnetization}(a), low-temperature magnetization shows metamagnetic behavior for $H \ || \ \langle 100 \rangle$ at high magnetic fields but not clearly for other directions. 
The metamagnetic behavior emerges in a low-temperature Fermi-liquid state and disappears at high temperatures above $\sim$12 K [$\sim T_{\chi}^{\mathrm{max}}$, the temperature of a local maximum in $\chi_{4f} (T)$].
The same temperature evolution of the metamagnetism was also observed in the parent compound YbRh$_2$Zn$_{20}$ \cite{Honda2013}.
However, the metamagnetic field $H_\textrm{m}$ slightly differs from that of the $x=0.014$ sample due to a smaller $T_{\mathrm{K}}(x=1.0) \sim 53.1$ K of the parent compound, which is estimated with $\gamma_{4f} = 740 \ \mathrm{mJ \ K^{-2} \ {(Yb \ mol)}^{-1}}$ \cite{Torikachvili2007} for $N = 8$.
In contrast, metamagnetism completely disappears in the solvent compound LuRh$_2$Zn$_{20}$ \cite{SM}. 
When extrapolating the metamagnetic field $H_\textrm{m}$ and amplitude of the metamagnetic anomaly (deviation from linearity) to $x \rightarrow 0$, intercepts of these quantities show a finite value.
Consequently, metamagnetic behavior in (Lu,Yb)Rh$_2$Zn$_{20}$ is a single-site effect in origin rather than a result of cooperative intersite phenomena. 
To characterize the metamagnetic behavior in the $x=0.014$ sample, we compare the data with the numerically calculated Coqblin-Schrieffer model for $N=8$ \cite{Hewson1983_2}. 
In Fig. \ref{LYRZ_Magnetization}(a), the qualitative feature of the metamagnetic behavior, except for the magnetic anisotropy, is reasonably reproduced with a theoretical curve (dashed line) for $T_1 = 17.7$ T, corresponding to $T_{\mathrm{K}} = 60.9$ K evaluated from $\gamma_{4f}$ \cite{T1}.
The observed difference and magnetic anisotropy are discussed in Sec. \ref{subsec_anisotropy_discussion}.

\subsection{Local maximum in magnetic susceptibility}
\label{subsec_chi}

An inflection point in $M(H)$ indicates a positive nonlinear term ($\propto H^3$) in the magnetization. As discussed in Refs. \cite{Hewson1993, Hewson1983_2}, the nonlinear term contributes to the third ingredient of the SU($N$) Kondo effect: a local maximum in $\chi(T)$.
The $\chi_{4f}(T)$ of the $x=0.014$ sample in Fig. \ref{LYRZ_macro_Tdep}(c) shows a local maximum at $T_{\chi}^{\mathrm{max}} \sim 12$ K for all $H$ directions that is qualitatively consistent with the numerical calculation of $N=8$ for $T_{\mathrm{K}} = 60.9$ K [light blue line in Fig. \ref{LYRZ_macro_Tdep}(c)] \cite{Rajan1983, Schlottmann1984_1, Schlottmann1984_2}. 
Notably, the cubic CEF effect cannot reproduce the local maximum in $\chi(T)$ in our analysis.

\subsection{Local minimum in quasielastic linewidth}
\label{subsec_quasielastic}

\begin{figure}[t]
	\includegraphics[keepaspectratio, width=8.6cm, clip]{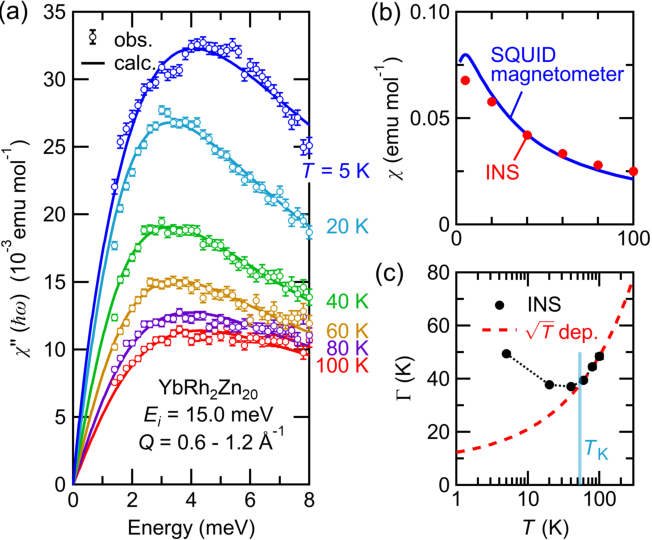}
	\caption{
	(a) Dynamical magnetic susceptibility $\chi''(\hbar\omega)$ integrated over $Q = 0.6-1.2 \ \text{\AA}^{-1}$. The solid lines are the results of fits with Eq. \eqref{Lorentzian}. (b) Temperature dependence of the static magnetic susceptibility determined using neutron-scattering measurements (red points) and magnetization measurements with a SQUID magnetometer (solid line). (c) Quasielastic linewidth $\Gamma(T)$ (black points) evaluated depending on the fit with Eq. \eqref{Lorentzian}. The black dotted line is a guide to the eye. The red dashed line shows the $\sqrt{T}$ dependence (dep.) expected for $N=2$. 
    \label{YRZ_INS}
	}
\end{figure}

The fourth ingredient of the SU($N$) Kondo effect, a local minimum in $\Gamma (T)$, is difficult to verify in the $x=0.014$ sample because the neutron-scattering signal is too weak to be observed in such a diluted system. Hence we examine the dynamical magnetic susceptibility of YbRh$_{2}$Zn$_{20}$.
We exhibit the imaginary part of the $Q$-integrated dynamical susceptibility of the parent compound YbRh$_{2}$Zn$_{20}$ in Fig. \ref{YRZ_INS}(a). The data set shows a typical temperature evolution for the quasielastic scattering resulting from the Kondo spin fluctuations, with no well-defined CEF excitations. To evaluate the linewidth of the quasielastic spectra $\Gamma$, the data were fitted with the following function \cite{Holland-Moritz1982}: 
\begin{equation}
	\chi''(Q,\hbar \omega, T) = f^2(Q) \chi(T) \hbar \omega \frac{\Gamma(T)}{(\hbar \omega)^2+\Gamma^2(T)}, 
	\label{Lorentzian}
\end{equation}
where $Q$ is the scattering vector, $\hbar \omega$ is the energy transfer, and $f(Q)$ is the magnetic form factor for \ce{Yb^{3+}} \cite{Boothroyd2020}.
The fitting parameters are the temperature-dependent static susceptibility $\chi(T)$ and quasielastic linewidth $\Gamma(T)$. 
The evaluated $\chi (T)$ and $\Gamma (T)$ are plotted in Figs. \ref{YRZ_INS}(b) and \ref{YRZ_INS}(c), respectively.
$\chi(T)$ is in good agreement with the magnetic susceptibility measured with a SQUID magnetometer \cite{Honda2013} within $\sim$20\% accuracy, reinforcing the reliability of the above analysis. 
Moreover, $\Gamma(T)$ shows a local minimum at approximately $40 \ \mathrm{K}$, which is very close to $T_\mathrm{K} = 53.1 \ \mathrm{K}$ of YbRh$_2$Zn$_{20}$, and roughly follows the $\sqrt{T}$ dependence above $T_{\mathrm{K}}$ as in Fig. \ref{YRZ_INS}(c) with the red dashed line.  
As predicted previously \cite{Bickers1985, Cox1985, Bickers1987}, such a local minimum in $\Gamma(T)$ possibly emerges for $N > 3$ as opposed to $N=2$ \cite{Murani1980, Horn1981, Walter1986}.
The local minimum of $\Gamma(T)$ reflects that the Kondo resonance peak for the multiorbital Kondo system is located away from the Fermi level, as mentioned by Schlottmann \cite{Schlottmann1989, Schlottmann1992}.

\subsection{Field-induced magnetic anisotropy}
\label{subsec_magnetic_anisotropy}

The above arguments provide evidence for the multiorbital Kondo ground state with SU(8) in the diluted Yb alloy (Lu$_{1-x}$Yb$_x$)Rh$_2$Zn$_{20}$.
However, a remarkable discrepancy from the SU(8) model is observed in the field evolution of magnetic anisotropy, as shown in Fig. \ref{LYRZ_Magnetization}(a).
For $x=0.014$, the characteristic crossover field (isotropic Kondo field, $H_\mathrm{K}$) is estimated to be $\mu_0 H_\mathrm{K} = k_\mathrm{B} T_\mathrm{K}/(g_J J \mu_\mathrm{B}) = 22.7 \ \mathrm{T}$ ($g_J$ is the Land\'{e} $g$ factor) \cite{Fenton1973, Ebihara2003}, above which the magnetic anisotropy becomes apparent.
Notably, the field-induced magnetic anisotropy emerges at a magnetic field significantly lower than $\mu_0 H_\mathrm{K}$.

\section{Discussion}

\subsection{Origin of field-induced magnetic anisotropy}
\label{subsec_anisotropy_discussion}
Here, we discuss field-induced single-ion magnetic anisotropy. 
Similar magnetization curves to those of the $x = 0.014$ sample were observed in the heavy-fermion paramagnet CeRu$_2$Si$_2$, which shows the metamagnetic behavior only for the easy axis $H \ || \ c$ \cite{Haen1987}.
However, the metamagnetic behavior in CeRu$_2$Si$_2$ is due to magnetic correlation \cite{Mignod1988, Flouquet2004} rather than the single-ion phenomenon. Since the orbital degeneracy in CeRu$_2$Si$_2$ is only $N = 2$ \cite{Willers2012}, the conduction electrons hybridize only with the $\Gamma_7$ CEF doublet ground state of Ce$^{3+}$ ions \cite{Haen1988}.  
The wave function of the $\Gamma_7$ doublet has $J_z = \ket{\pm 5/2}$ and $\ket{\mp 3/2}$ components, and microscopic experiments in CeRu$_2$Si$_2$ revealed $\ket{\pm 5/2}$ as the main $J_z$ component \cite{Boucherle2001, Willers2012}.
Thus the ground-state wave function with a uniaxial shape leads to strong magnetic anisotropy \cite{Haen1987}.

In contrast, as mentioned, metamagnetic behavior in the $x=0.014$ sample is a single-site effect in origin. Moreover, $M(H)$ is explained by the SU(8) Kondo model, except for the magnetic anisotropy. 
As expected, no magnetic anisotropy emerges in the SU($N$) Kondo ground state even in a finite magnetic field because anisotropic interactions are neglected in the SU($N$) Kondo (Coqblin-Schrieffer) model. 
As intersite magnetic interactions between the magnetic Yb ions are negligible in the $x=0.014$ sample, the CEF effect causes the single-ion magnetic anisotropy. The cubic CEF effect lifts the eightfold multiplet of the Yb ion into two doublets ($\Gamma_6$ and $\Gamma_7$) and a quartet ($\Gamma_8$). 
However, the energy gap among these CEF states is small owing to a nearly spherical potential in the Frank-Kasper cage. In addition, the magnetic anisotropy is further reduced in combination with the relatively large $c$-$f$ hybridization effect of $T_{\mathrm{K}} = 60.9$ K in the region for low $T$ and low $H$ (strong-coupling regime). 
Small nonzero magnetic anisotropy was observed in the $x=0.014$ sample, as shown in Fig. \ref{LYRZ_macro_Tdep}(c). The magnetic anisotropy ratio is $M_{100}/M_{111} = 1.08$ between the easy and hard axes at 1.8 K for $\mu_0H = 2$ T, reflecting a weak CEF effect.

The anisotropy ratio at $\mu_0 H = 14.5$ T is enhanced to $M_{100}/M_{111} = 1.37$, which is 1.27 times larger than that at 2 T but is still weak compared with CeRu$_2$Si$_2$. Such field-induced single-ion magnetic anisotropy could be interpreted as a modified ground-state wave function. 
Although the Zeeman effect slightly lifts the CEF levels at low magnetic fields, conduction electrons screen the localized moment and form a Kondo cloud around the Yb ion. 
This low-field Kondo cloud is almost isotropic because all eight wave functions are mixed through the multiorbital Kondo effect, resulting in nearly isotropic magnetic behavior.
As the magnetic field increases, the Zeeman effect further lifts the CEF levels. 
The characteristic energy scale $\Delta_f(H)$, defined as the difference between the highest and lowest CEF levels, eventually exceeds the energy scale of the Kondo temperature, i.e., $\Delta_f(H) > k_\mathrm{B} T_\mathrm{K}$, as schematically illustrated in Fig. \ref{LYRZ_Magnetization}(b). 
At such high magnetic fields, $f$ electrons can no longer occupy the excited CEF levels exceeding $k_\mathrm{B} T_\mathrm{K}$ through the multiorbital Kondo effect. Thus magnetic fields gradually deform the ground-state wave function, contributing to magnetic anisotropy; therefore the field-induced magnetic anisotropy of (Lu$_{1-x}$Yb$_x$)Rh$_2$Zn$_{20}$ gradually appears. Based on the above discussion, the magnetic anisotropy should appear near the magnetic field where $\Delta_f(H)$ equals $k_\mathrm{B} T_\mathrm{K}$.
In $f$-electron compounds, owing to the energy level splitting at 0 T due to the CEF effect, $\Delta_f(H)$ reaches $k_\mathrm{B}T_\mathrm{K}$ with a small Zeeman splitting compared with no level splitting at 0 T.
Namely, a weak CEF splitting may result in field-induced magnetic anisotropy at lower magnetic fields. 

\subsection{CEF energy scale}

Finally, we discuss the CEF energy scale $\Delta_\mathrm{CEF}$; i.e., $\Delta_f(H = 0)$ for $x = 0.014$.
As mentioned above, the upper limit of $\Delta_\mathrm{CEF}/k_\mathrm{B}$ can be $\sim$60 K because $\Delta_\mathrm{CEF}/k_\mathrm{B} < T_\mathrm{K} = 60.9 \ \mathrm{K}$. 
Furthermore, the lower limit can be deduced from the above discussion. 
Since the magnetic anisotropy appears at $\sim$5 T for $x = 0.014$, the formula $\Delta_f(\mu_0 H = 5 \ \mathrm{T})/k_\mathrm{B} \sim  T_\mathrm{K} = 60.9 \ \mathrm{K}$ can be derived.
As the maximum and minimum expectation values of $J_z$ for the Yb$^{3+}$ ion are $7/2$ and $-7/2$, respectively, the increment of $\Delta_f(H)/k_\mathrm{B}$ due to the Zeeman effect is at most $\{g_J \mu_\mathrm{B} \cdot 7/2 - g_J \mu_\mathrm{B} \cdot (-7/2)\}/k_\mathrm{B} = 5.4 \ \mathrm{K}$ per 1 T.
Thus $\{ \Delta_f(\mu_0 H = 5 \ \mathrm{T}) - \Delta_f(H = 0) \}/k_\mathrm{B} \le 27.0 \ \mathrm{K}$.
By combining this inequality with $\Delta_f(\mu_0 H = 5 \ \mathrm{T})/k_\mathrm{B} \sim 60.9 \ \mathrm{K}$, the lower limit of $\Delta_\mathrm{CEF}/k_\mathrm{B}$ can be estimated to be 33.9 K.
The small CEF energy scale of $30 \ \mathrm{K} \lesssim \Delta_\mathrm{CEF}/k_\mathrm{B} \lesssim 60 \ \mathrm{K}$ in (Lu$_{1-x}$Yb$_x$)Rh$_2$Zn$_{20}$ can be possible because $\Delta_\mathrm{CEF}/k_\mathrm{B}$ for the isovalent compound YbCo$_2$Zn$_{20}$ was evaluated to be $\sim$30 K \cite{Takeuchi2011, Kaneko2012}. 
To clarify the CEF energy scheme, we are currently investigating high-magnetic-field experiments in the diluted Yb system of (Lu$_{1-x}$Yb$_x$)Rh$_2$Zn$_{20}$ and performing a quantitative comparison with the calculation of the multiorbital Kondo model considering the CEF and Zeeman effect.

\section{Conclusion}
\color{black}

We examined the field-induced single-ion magnetic anisotropy and ground-state properties of (Lu$_{1-x}$Yb$_{x}$)Rh$_2$Zn$_{20}$.
Specific heat, magnetization, resistivity, and neutron-scattering measurements revealed that the ground-state properties, except for the field-induced magnetic anisotropy, could be qualitatively reproduced by the SU($N = 8$) Kondo model with $T_\mathrm{K} = 60.9 \ \mathrm{K}$.
Moreover, the magnetic anisotropy becomes apparent only at approximately 5 T, which is a much lower magnetic field than the isotropic Kondo crossover field of 22.7 T.
The apparent lowering of the Kondo crossover field could result from the blurred CEF effect on the Yb ion.
Our results suggest that the diluted Yb alloy, (Lu$_{1-x}$Yb$_x$)Rh$_2$Zn$_{20}$, is an appealing material in understanding the magnetic anisotropy in multiorbital Kondo physics.

\begin{acknowledgments}
We would like to thank K. Kaneko, T. Osakabe, K. Miyake, and K. Iwasa for our fruitful discussions.
A major part of this work was supported by JSPS KAKENHI Grant No. JP19K03737, No. JP19H05164, and No. JP21H00139. 
We thank Dr. Sakamoto from Analytical Research Core for Advanced Materials (chemical composition analysis by ICP-AES) and Laboratory of Low Temperature Material Science (magnetometry with a SQUID magnetometer), Institute for Materials Research, Tohoku University, for the experimental support provided. 
A part of this work was supported by JSPS KAKENHI (Grant No. JP16H02125, No. 20K03851, No. JP21H03732, No. JP21H04448, No. JP21H04987, No. JP22H05145, No. 23H01132, and No. 23K03314), JST FOREST (Grant No. JPMJFR202V), JST SPRING (Grant No. JPMJSP2114), and the REIMEI Research Program of JAEA. 
The neutron-scattering experiments were performed under the J-PARC user program (Proposal No. 2019B0218: BL01 4SEASONS at MLF). 
A part of the macroscopic measurements was supported by the Visiting Researchers Program of the Institute of Solid State Physics, The University of Tokyo. 
\end{acknowledgments}

\bibliography{bibliography}

\begin{thebibliography}{74}
\bibitem [{met()}]{metamag}%
  \BibitemOpen
  \href@noop {} {}\bibinfo {note} {Ordinary metamagnetic behavior may be
  observed in itinerant paramagnets and antiferromagnetically ordered materials
  when the Zeeman energy overcomes a potential barrier resulting from intersite
  magnetic interactions between magnetic ions. Such ordinary metamagnetism is a
  cooperative phenomenon that disappears with decreasing magnetic ion
  concentrations in a material, as illustrated in Fig.
  \ref{magnetization_illustration}(a). In contrast, the Kondo effect is a
  single-site phenomenon resulting from the spin-flip interaction between the
  localized and conduction electrons close to the Fermi surface. Therefore the
  metamagnetism originating from the multiorbital Kondo effect appears even in
  diluted limits, as shown in Fig.
  \ref{magnetization_illustration}(b).}\BibitemShut {Stop}%
\bibitem [{WR()}]{WR}%
  \BibitemOpen
  \href@noop {} {}\bibinfo {note} {The relation $R_\mathrm{W} = N/(N-1)$ is
  valid for a Kondo regime where the number of 4$f$ electrons is close to an
  integer and then charge fluctuations are negligible. Therefore localized
  magnetic behavior (Curie-Weiss law) may be observed at high temperatures as
  being consistent with our result of $\chi_{4f}^{-1}(T)$ in Fig.
  \ref{LYRZ_macro_Tdep}(b).}\BibitemShut {Stop}%
\bibitem [{TK()}]{TK}%
  \BibitemOpen
  \href@noop {} {}\bibinfo {note} {$T_{\mathrm{K}}$ was evaluated with the
  following relation: $T_{\mathrm{K}} = W \times T_0 = W \times (N-1)\pi
  R/(6\gamma_{4f})$ \cite{Rajan1983, Kaihe2005}. Here, $W$ is the Wilson number
  1.290 \cite{Kaihe2005}, $R$ is the gas constant, $N$ is the total degeneracy,
  and $T_0$ is a measure of the width of the Abrikosov-Suhl (Kondo) resonance
  peak.}\BibitemShut {Stop}%
\bibitem [{T1()}]{T1}%
  \BibitemOpen
  \href@noop {} {}\bibinfo {note} {$T_1$ can be written as $24\pi C_1 T_0 /
  N(N^2-1)$, where $C_1$ is a scaling constant in proportion to $J$ ($1.686$
  for $J=7/2$) and $T_0$ is a characteristic temperature corresponding to the
  Kondo temperature as described in Ref. \cite{TK}.}\BibitemShut {Stop}%
\end{thebibliography}
\nocite{*}

\end{document}



\title{Supplementary Material: Observation of Field-Induced Single-Ion Magnetic Anisotropy in a Multiorbital Kondo Alloy (Lu,Yb)Rh$_2$Zn$_{20}$}

\author{T. Kitazawa}
\email[takafumi.kitazawa.s5@dc.tohoku.ac.jp]{}
\affiliation{Graduate School of Science, Tohoku University, Sendai 980-8578, Japan}
\affiliation{Advanced Science Research Center, Japan Atomic Energy Agency, Tokai, Ibaraki 319-1195, Japan}

\author{Y. Ikeda}
\affiliation{Institute for Materials Research, Tohoku University, Sendai 980-8577, Japan}
\author{T. Sakakibara}
\affiliation{The Institute for Solid State Physics, The University of Tokyo, Kashiwa, Chiba 277-8581, Japan}
\author{A. Matsuo}
\affiliation{The Institute for Solid State Physics, The University of Tokyo, Kashiwa, Chiba 277-8581, Japan}
\author{Y. Shimizu}
\affiliation{Institute for Materials Research, Tohoku University, Oarai, Ibaraki 311-1313, Japan}
\author{Y. Tokunaga}
\affiliation{Advanced Science Research Center, Japan Atomic Energy Agency, Tokai, Ibaraki 319-1195, Japan}
\author{\\Y. Haga}
\affiliation{Advanced Science Research Center, Japan Atomic Energy Agency, Tokai, Ibaraki 319-1195, Japan}
\author{K. Kindo}
\affiliation{The Institute for Solid State Physics, The University of Tokyo, Kashiwa, Chiba 277-8581, Japan}
\author{Y. Nambu}
\affiliation{Institute for Materials Research, Tohoku University, Sendai 980-8577, Japan}
\affiliation{Fusion Oriented Research for Disruptive Science and Technology, Japan Science and Technology Agency, Kawaguchi, Saitama 332-0012, Japan}
\affiliation{Organization for Advanced Studies, Tohoku University, Sendai 980-8577, Japan}
\author{K. Ikeuchi}
\affiliation{Neutron Science and Technology Center, Comprehensive Research Organization for Science and Society (CROSS), Tokai, Ibaraki 319-1106, Japan}
\author{K. Kamazawa}
\affiliation{Neutron Science and Technology Center, Comprehensive Research Organization for Science and Society (CROSS), Tokai, Ibaraki 319-1106, Japan}
\author{M. Ohkawara}
\affiliation{Institute for Materials Research, Tohoku University, Sendai 980-8577, Japan}
\author{M. Fujita}
\affiliation{Institute for Materials Research, Tohoku University, Sendai 980-8577, Japan}




\maketitle




%

\section{Composition Analysis}

Table \ref{ICP} summarizes the composition analysis results obtained via inductively coupled plasma atomic emission spectroscopy (ICP-AES) for samples with nominal Yb concentrations of $X = 0$ and 0.1.
The composition was normalized such that the sum of Yb and Lu concentration equals 1. Thus, the nominal ratio of (Lu + Yb), Rh, and Zn is almost identical to the measured composition ratio. 
The evaluated Yb concentration in $X = 0$ samples (i.e., impurity in a non-magnetic reference compound LuRh$_2$Zn$_{20}$) is at most $4.9 \times 10^{-4}$ ($< 490$ ppm). In contrast, the Yb concentration in $X = 0.1$ samples was evaluated as 0.0137(6), approximately 1/10 of the nominal concentration. The evaluated Yb concentration of $\sim$1.4\% is sufficiently smaller than that of the percolation limit for a cubic lattice, suggesting that magnetic interactions find it challenging to keep a long-range correlation. Furthermore, the probability that two Yb ions occupy the nearest neighbor sites is approximately 5\% (i.e., 95\% Yb ions are isolated) for $x = 0.014$ according to the binomial distribution. This fact also supports the observed phenomena in (Lu,Yb)Rh$_2$Zn$_{20}$ of single-site origin.

\begin{table}[h]
\centering
\caption{\label{ICP}
Inductively coupled plasma atomic emission spectroscopy analysis results for (Lu,Yb)Rh$_2$Zn$_{20}$. $X$ and $x$ denote the nominal and measured Yb concentration, respectively.}
\catcode`!=\active \def!{\phantom{.}}
\catcode`?=\active \def?{\phantom{0}}
\vspace{2pt}
\begin{minipage}[t]{0.6\textwidth}
\begin{ruledtabular}
\begin{tabular}{cccccc}
$X$ &
&
\begin{tabular}{c} Yb \\ ($x$) \end{tabular}&
\begin{tabular}{c} Lu \\ ($1-x$) \end{tabular}&
Rh &
Zn \\
\midrule
0 & sample 1 & $< 4.9 \times 10^{-4}$ \footnotemark[1] & 1 & 1.91 & 19.84 \\
& sample 2 & $< 4.9 \times 10^{-4}$ \footnotemark[1] & 1 & 1.95 & 19.93 \\ \cmidrule{2-6}
& average & & 1 & 1.93 & 19.88 \\ \midrule
0.1 & sample 1 & 0.0143 & 0.9857 & 1.97 & 19.89 \\
& sample 2 & 0.0131 & 0.9869 & 1.96 & 19.87 \\ \cmidrule{2-6}
& average & 0.0137 & 0.9863 & 1.96 & 19.88 \\
\end{tabular}
\end{ruledtabular}
\footnotetext[1]{limit of the quantitative analysis}
\end{minipage}
\end{table}

\section{Specific heat}

\begin{figure}[t]
	\includegraphics[keepaspectratio, width=0.37\textwidth, clip]{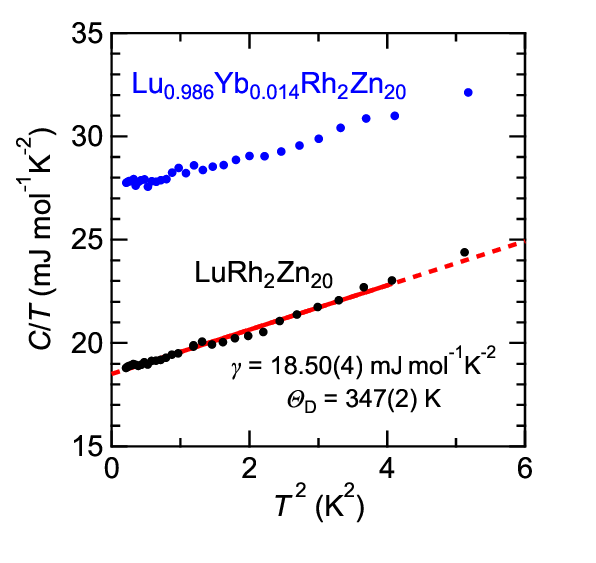}
	\caption{Specific heat divided by temperature $C/T$ versus $T^2$ for Lu$_{0.986}$Yb$_{0.014}$Rh$_2$Zn$_{20}$ and LuRh$_2$Zn$_{20}$. The solid line is the best fit with $C/T = \gamma + \beta T^2$ to the experimental data below $5 \ \mathrm{K}^2$ of LuRh$_2$Zn$_{20}$ (black closed circle). The dashed line is the extended solid line. \label{C_over_T_sample}}
\end{figure}

Figure \ref{C_over_T_sample} shows the $T^2$ dependence of the specific heat divided by temperatures below $6 \ \mathrm{K}^2$ for $x = 0$ (black points) and 0.014 (blue points) samples. The specific heat shows the Fermi liquid behavior in addition to a typical phonon contribution, i.e., $C = \gamma T + \beta T^3$, in both samples. Assuming that the lattice specific heat for $x = 0.014$ samples is the same as that for $x = 0$, we extracted the $4f$-electron contribution $C_{4f}$, shown in Fig. 2(a) in the manuscript, as $C_{4f} = C(x = 0.014) - C(x = 0)$.

The electronic specific heat coefficient $\gamma$ and phonon-term coefficient $\beta$ for LuRh$_2$Zn$_{20}$ are evaluated by linearly fitting the data below $T^2 < 5 \ \mathrm{K}^2$ to $18.50(4) \ \mathrm{mJ \ mol}^{-1} \ \mathrm{K}^{-2}$ and $1.07(2) \ \mathrm{mJ} \ \mathrm{mol}^{-1} \ \mathrm{K}^{-4}$, respectively.
Thus, the Debye temperature is $\varTheta_\mathrm{D} = 347(2) \ \mathrm{K}$.

\section{Magnetization}

\subsection{Magnetization curve in a low magnetic field}

\begin{figure}[t!]
	\includegraphics[keepaspectratio, width=0.87\textwidth, clip]{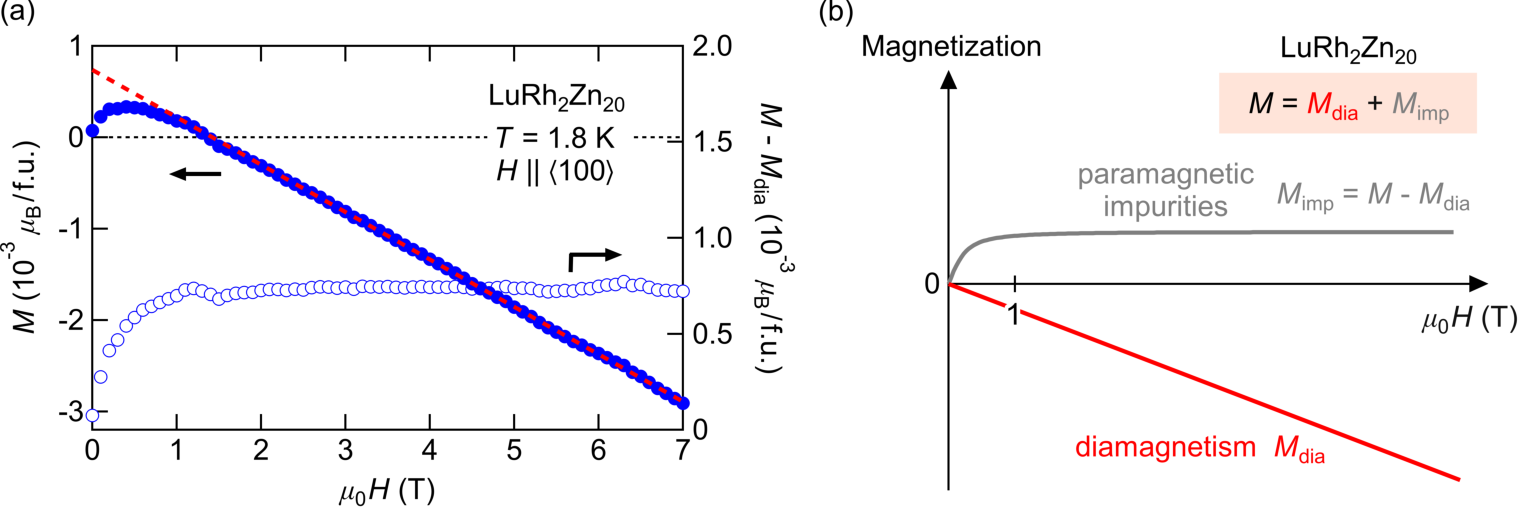}
	\caption{Magnetic field dependence of magnetization for LuRh$_2$Zn$_{20}$. (a) Observed magnetization curve $M(H)$ (left axis) and the magnetization after subtracting the diamagnetic contribution from $M(H)$ (right axis) along the $\langle 100 \rangle$ direction at 1.8 K. (b) Schematic of the diamagnetic $M_\mathrm{dia}(H)$ (red line) and paramagnetic impurities contribution $M_\mathrm{imp}(H)$ (gray line) in LuRh$_2$Zn$_{20}$. \label{LRZ_magnetization}}
\end{figure}

\begin{table}[t!]
\centering
\caption{\label{chi_dia}
 Magnetic susceptibility of the diamagnetic component in LuRh$_2$Zn$_{20}$, $\chi_\mathrm{dia} = M_\mathrm{dia}/H$, evaluated from the magnetization curve of LuRh$_2$Zn$_{20}$ at 1.8 K.}
\catcode`!=\active \def!{\phantom{.}}
\catcode`?=\active \def?{\phantom{0}}
\vspace{2pt}
\begin{minipage}[t]{0.3\textwidth}
\begin{ruledtabular}
\begin{tabular}{rc}
&
$\chi_\mathrm{dia} \ [10^{-4} \ \mathrm{emu \ mol}^{-1}]$ \\
\midrule
$H \ || \ \langle 100 \rangle$ & -2.898(6) \\
$\langle 110 \rangle$ & -2.882(2) \\
$\langle 111 \rangle$ & -2.890(3) \\
\end{tabular}
\end{ruledtabular}
\end{minipage}
\end{table}

Figure \ref{LRZ_magnetization}(a) displays the magnetization curves of \ce{LuRh2Zn20} for $H \ || \ \langle 100 \rangle$.
\ce{LuRh2Zn20} exhibits diamagnetic behavior above 1 T. The slight increase in magnetization below 0.5 T is due to paramagnetic impurities.
Therefore, the magnetization of \ce{LuRh2Zn20} can be decomposed into two components: (i) diamagnetic $M_\mathrm{dia}$ and (ii) paramagnetic $M_\mathrm{imp}$, as schematically illustrated in Fig. \ref{LRZ_magnetization}(b).
Similar behavior was observed in the $\langle 110 \rangle$ and $\langle 111 \rangle$ directions.
To extract the diamagnetic components $M_\mathrm{dia}$, we linearly fit the magnetization data at 1.8 K in the range $2 \ \mathrm{T} \le \mu_0 H \le 7 \ \mathrm{T}$ for each direction.
The slopes of the fitting lines $\chi_\mathrm{dia}$ are listed in Table \ref{chi_dia}. As shown in the open circles of Fig. \ref{LRZ_magnetization}(a), the magnetization curve at 1.8 K subtracted by $M_\mathrm{dia}$ ($= \chi_\mathrm{dia} H$) represents the paramagnetic behavior.

\begin{figure}[t!]
	\includegraphics[keepaspectratio, width=0.96\textwidth, clip]{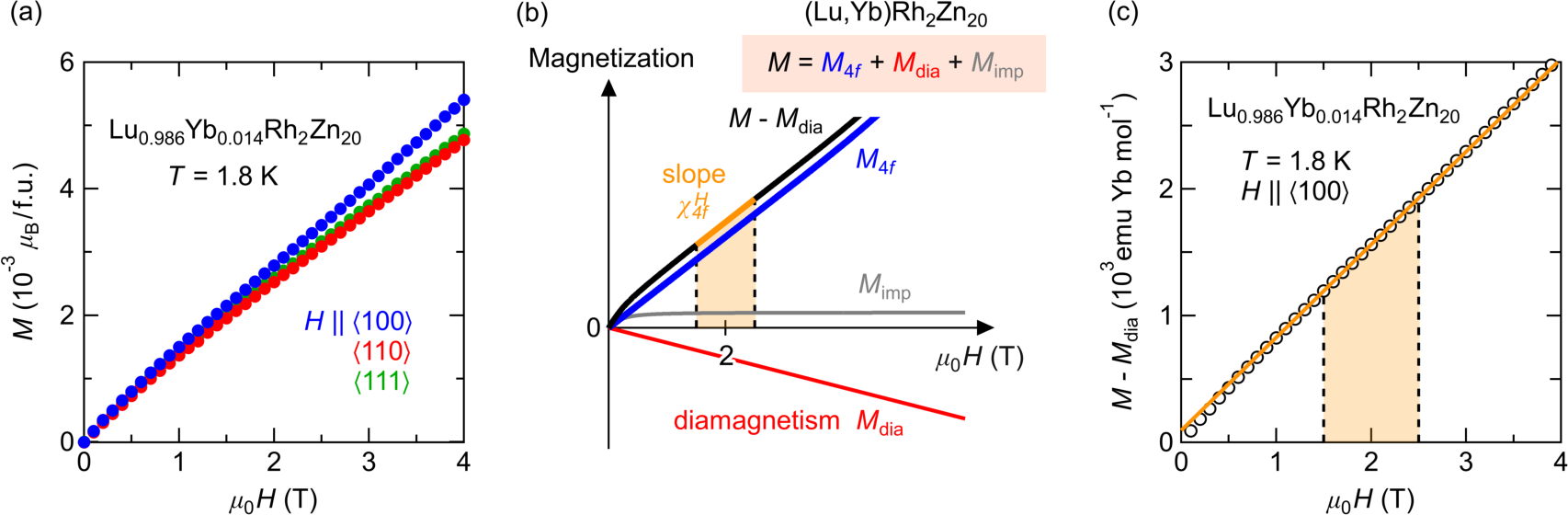}
	\caption{Magnetic field dependence of magnetization for Lu$_{0.986}$Yb$_{0.014}$Rh$_2$Zn$_{20}$. (a) Observed magnetization curves $M(H)$ along the $\langle 100 \rangle$, $\langle 110 \rangle$, and $\langle 111 \rangle$ directions at 1.8 K. (b) Schematic of the diamagnetic $M_\mathrm{dia}(H)$ (red line), paramagnetic $M_\mathrm{imp}(H)$ (gray line), and $4f$-electrons contribution $M_{4f}$ (blue line) in (Lu,Yb)Rh$_2$Zn$_{20}$. The slope of the yellow line around 2 T on the $M - M_\mathrm{dia}$ data is considered as the $4f$-electron magnetic susceptibility $\chi_{4f}^H$. (c) $M - M_\mathrm{dia}$ for the $\langle 100 \rangle$ direction at 1.8 K. The dark yellow line is the best fit to $M - M_\mathrm{dia}$ data (open circle) in the range $1.5 \ \mathrm{T} \le \mu_0 H \le 2.5 \ \mathrm{T}$. The slope of the line, $\chi_{4f}^H (1.8 \ \mathrm{K})$, is listed in Table \ref{chi_LYRZ}. \label{LYRZ_magnetization}}
\end{figure}

\begin{table}[t!]
\centering
\caption{\label{chi_LYRZ} 
Temperature-independent magnetic susceptibility, $\chi_0$, evaluated using the modified Curie--Weiss fit, magnetic susceptibility at $1.8 \ \mathrm{K}$, $\chi_{4f}(1.8 \ \mathrm{K})$ and $\chi_{4f}^H(1.8 \ \mathrm{K})$, in Lu$_{0.986}$Yb$_{0.014}$Rh$_2$Zn$_{20}$.
All values are in the unit of $\mathrm{10^{-3}emu \ Yb\text{-}mol^{-1}}$.}
\catcode`!=\active \def!{\phantom{.}}
\catcode`?=\active \def?{\phantom{0}}
\vspace{2pt}
\begin{minipage}[t]{0.47\textwidth}
\begin{ruledtabular}
\begin{tabular}{rccc}
&
$\chi_0$ &
$\chi_{4f}(1.8 \ \mathrm{K})$ &
$\chi_{4f}^H(1.8 \ \mathrm{K})$ \\
\midrule
$H \ || \ \langle 100 \rangle$ & -6.76(2)? & 73.1(1) & 73.33(3) \\
$\langle 110 \rangle$ & -5.88(7)? & 66.5(2) & 67.2(2)? \\
$\langle 111 \rangle$ & -5.04(12) & 67.6(6) & 68.3(1)? \\
\end{tabular}
\end{ruledtabular}
\end{minipage}
\end{table}

Next, we show the magnetization curves for the $x = 0.014$ sample in Fig. \ref{LYRZ_magnetization}(a).
The slightly convex curve below 1 T is due to paramagnetic impurities, which almost saturate above 1 T, as observed in LuRh$_2$Zn$_{20}$. Therefore, the linear increase in $1 \ \mathrm{T} \le \mu_0 H \le 3 \ \mathrm{T}$ can be ascribed to the magnetization contribution from the diluted Yb $4f$-electrons. Thus, \ce{(Lu, Yb)Rh2Zn20} has three components, as illustrated in Fig. \ref{LYRZ_magnetization}(b): (i) diamagnetic contribution from LuRh$_2$Zn$_{20}$ $M_\mathrm{dia}$, (ii) paramagnetic impurities $M_\mathrm{imp}$, and (iii) Yb $4f$-electrons $M_{4f}$.
The Yb-$4f$ magnetic susceptibility at 1.8 K, $\chi_{4f}^{H}(1.8 \ \mathrm{K})$, can be evaluated from the slope of $M - M_\mathrm{dia}$ at 1.8 K at around 2 T, in which the impurity contribution is considered to be saturated, as shown in Fig. \ref{LYRZ_magnetization}(b). The dark yellow line in Fig. \ref{LYRZ_magnetization}(c) is the linear fit in the range $1.5 \ \mathrm{T} \le \mu_0 H \le 2.5 \ \mathrm{T}$. The evaluated values for each field direction are listed in Table \ref{chi_LYRZ}.

\subsection{Temperature evolution of magnetization}

\begin{figure}[t!]
	\includegraphics[keepaspectratio, width=0.37\textwidth, clip]{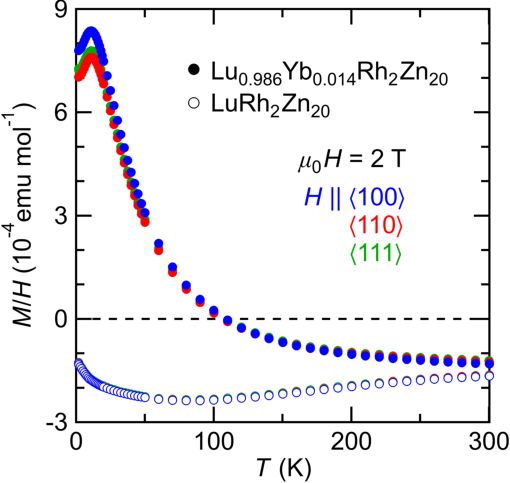}
	\caption{Magnetization divided by magnetic field ($\mu_0 H = 2 \ \mathrm{T}$) versus temperature for Lu$_{0.986}$Yb$_{0.014}$Rh$_2$Zn$_{20}$ (filled circle) and \ce{LuRh2Zn20} (open circle). \label{M_over_H_raw}}
    \vspace{5mm}
	\includegraphics[keepaspectratio, width=0.9\textwidth, clip]{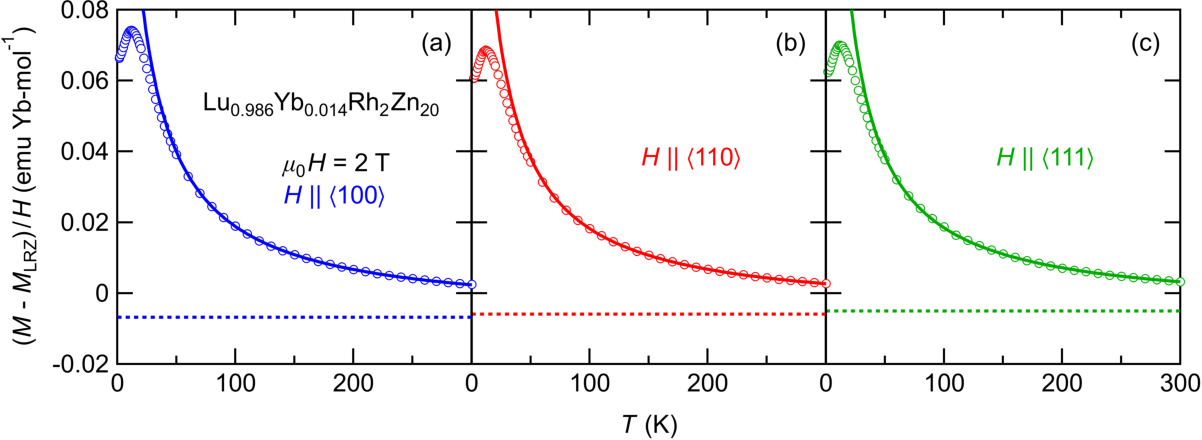}
	\caption{Temperature dependence of $(M - M_\mathrm{LRZ})/H$ along the (a) $\langle 100 \rangle$, (b) $\langle 110 \rangle$, and (c) $\langle 111 \rangle$ directions at 2 T. Solid lines represent the modified Curie--Weiss fit in the temperature range of $100 \ \mathrm{K} \le T \le 300 \ \mathrm{K}$, and the dotted lines represent the temperature-independent magnetic susceptibility $\chi_0$, whose values are listed in Table \ref{chi_LYRZ}.\label{M_over_H_subt_LRZ}}
\end{figure}

The temperature dependence of magnetization divided by the magnetic field at 2 T for $x = 0$ and 0.014 samples is displayed in Fig. \ref{M_over_H_raw}. $M/H$ of \ce{LuRh2Zn20} shows an upturn ascribed to paramagnetic impurities and a slight increase above 100 K. The later temperature dependence might be due to the orbital diamagnetism, as observed in LaV$_{2-x}$Ti$_x$Al$_{20}$ ($0 \le x \le 0.2$) \cite{Hirose2015}.
By substituting only 1.4 at$\%$ Yb to \ce{LuRh2Zn20}, a local maximum appears at 11 K for each magnetic field direction.
As LuRh$_2$Zn$_{20}$ does not show such a local maximum, this anomaly in (Lu,Yb)Rh$_2$Zn$_{20}$ can be interpreted as the Yb single-site phenomena.

To extract the $4f$-electron contribution of magnetic susceptibility, $\chi_{4f}(T)$, for $x = 0.014$ sample, we subtracted $M/H$ of $x = 0$ sample (hereinafter referred to as $M_\mathrm{LRZ}/H$) from that of $x = 0.014$ sample.
The data of $(M - M_\mathrm{LRZ})/H$, shown in Fig. \ref{M_over_H_subt_LRZ}, were fitted by the modified Curie--Weiss law $(M - M_\mathrm{LRZ})/H = C/(T - \theta_\mathrm{p}) + \chi_0$ in the temperature range of 100 to 300 K. Here $C$ is the Curie constant, $\theta_\mathrm{p}$ is the Curie--Weiss temperature, and $\chi_0$ is the temperature-independent magnetic susceptibility.
The fitting curves for each magnetic direction are shown with solid lines in Fig. \ref{M_over_H_subt_LRZ}.
The fitting parameters $\theta_\mathrm{p}$ and $\chi_0$ are listed in Tables I and \ref{chi_LYRZ}, respectively. The effective magnetic moment $\mu_\mathrm{eff}$ derived from $C$ is listed in Table I.
Finally, the Yb $4f$-electron component $M_{4f}/H$ can be evaluated from $M_{4f}/H = (M - M_\mathrm{BG})/H$.
Here, $M_\mathrm{BG}$ is the sum of $M_\mathrm{LRZ}$ and $\chi_0 H$ at 2 T.
The values of $\chi_{4f} = M_{4f}/H$ at $1.8 \ \mathrm{K}$, $\chi_{4f}(1.8 \ \mathrm{K})$, for the $\langle 100 \rangle$, $\langle 110 \rangle$, and $\langle 111 \rangle$ directions are listed in Tables I and \ref{chi_LYRZ}. $\chi_{4f}(1.8 \ \mathrm{K})$ for each direction is close to $\chi_{4f}^H(1.8 \ \mathrm{K})$.

\subsection{Yb concentration for \boldmath$X = 0.25$}
\label{estimation_0p25}

\begin{figure}[t!]
	\includegraphics[keepaspectratio, width=0.37\textwidth, clip]{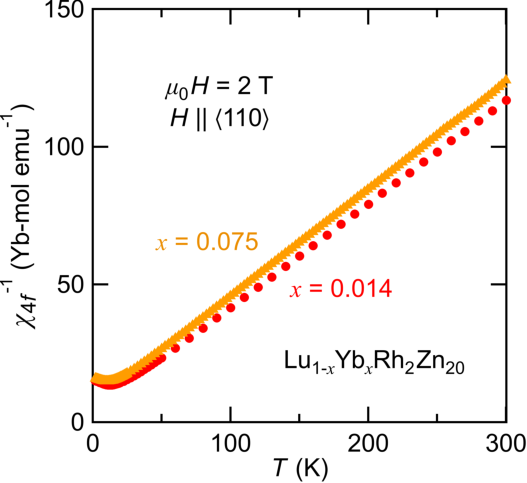}
	\caption{Temperature dependence of reciprocal magnetic susceptibility for $H \ || \ \langle 110 \rangle$ in Lu$_{1-x}$Yb$_{x}$Rh$_2$Zn$_{20}$ $(x = 0.014, 0.075)$ \label{LYRZ_MT_0p075}}
\end{figure}

\begin{table}[t!]
\centering
\caption{\label{LYRZ_parameters} 
Actual Yb concentration $x$, effective magnetic moment $\mu_\mathrm{eff}$, Curie--Weiss temperature $\theta_\mathrm{p}$, magnetic susceptibility $\chi_{4f}$ at $1.8 \ \mathrm{K}$, and temperature at which $\chi_{4f}(T)$ exhibits a local maximum, $T_{\chi}^{ \mathrm{max}}$, for $H \ || \ \langle 110 \rangle$ in the samples of nominal Yb concentration $X = 0.1$ and 0.25.} 
\catcode`!=\active \def!{\phantom{.}}
\catcode`?=\active \def?{\phantom{0}}
\vspace{2pt}
\begin{minipage}[t]{0.57\textwidth}
\begin{ruledtabular}
\begin{tabular}{cccccc}
$X$ &
$x$ &
\begin{tabular}{c} $\mu_\mathrm{eff}$ \\ $[\mu_\mathrm{B}/\mathrm{Yb}]$ \end{tabular}&
\begin{tabular}{c} $\theta_\mathrm{p}$ \\ $[\mathrm{K}]$ \end{tabular}&
$\chi_{4f}(1.8 \ \mathrm{K})$\footnotemark[1]&
\begin{tabular}{c} $T_{\chi}^{\mathrm{max}}$ \\ $[\mathrm{K}]$ \end{tabular}
\\
\midrule
0.1? & 0.014 & 4.61(3) & -10.2(9) & 66.5(2) & 12!? \\
0.25 & 0.075 & 4.54? & -17.9(3) & 63.1(3) & 10.5 \\
\end{tabular}
\end{ruledtabular}
\footnotetext[1]{The unit is $\mathrm{10^{-3}emu \ Yb\text{-}mol^{-1}}$.}
\end{minipage}
\end{table}

As mentioned in Section II of the main manuscript, while the actual Yb concentration $x = 0.014$ in $X = 0.1$ sample was examined using the ICP-AES analysis, the Yb concentration for $X = 0.25$ was evaluated using $M/H(T)$.
Since $\mu_\mathrm{eff}$ derived from the modified Curie--Weiss law in the range of $100 \ \mathrm{K} \le T \le 300 \ \mathrm{K}$ for $x = 0.014$ sample is close to $4.54 \ \mu_{\mathrm{B}}$/Yb expected for the free Yb$^{3+}$ ion (see Table I), we estimate the Yb concentration of $X = 0.25$ sample by assuming $\mu_\mathrm{eff} = 4.54 \ \mu_{\mathrm{B}}$/Yb.
By determining Yb concentration from $M/H(T)$ in the temperature range $100 \ \mathrm{K} \le T \le 300 \ \mathrm{K}$ so that $\mu_\mathrm{eff} = 4.54 \ \mu_{\mathrm{B}}$/Yb, we obtained $x = 0.075$.
$\chi_{4f}(T)$ for $x = 0.014$ and 0.075 is shown in Fig. \ref{LYRZ_MT_0p075}; the evaluated parameters are summarized in Table \ref{LYRZ_parameters}.
$\chi_{4f}(1.8 \ \mathrm{K})$, which approximates the zero-temperature magnetic susceptibility $\chi_{4f}(0)$ (see Fig. 2(c)), for $x = 0.014$ is almost the same as that for $x = 0.075$, as listed in Table \ref{LYRZ_parameters}.
Thus, the Kondo temperature $T_\mathrm{K}$ for $x = 0.075$ is inferred to be $\sim$ 60 K since $\chi_{4f}(0)$ is proportional to $T_\mathrm{K}^{-1}$.
$x=0.075$ sample was used in the resistivity measurements shown in Figs. 2(d) and \ref{LYRZ_rho_Tdep}.

\section{Electrical resistivity}

\begin{figure}[t]
	\includegraphics[keepaspectratio, width=0.43\textwidth, clip]{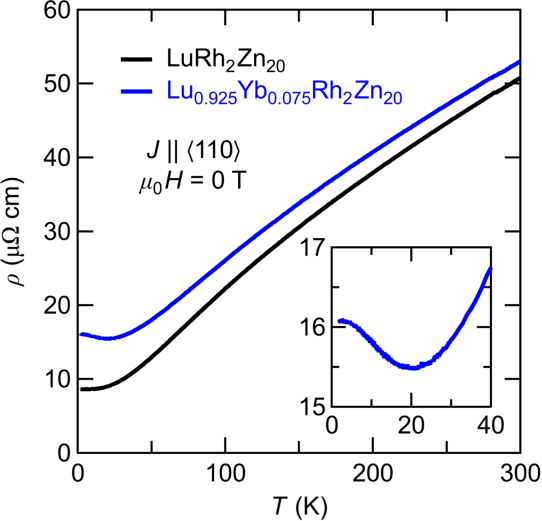}
	\caption{Temperature dependence of resistivity for Lu$_{0.925}$Yb$_{0.075}$Rh$_2$Zn$_{20}$ and LuRh$_2$Zn$_{20}$. The inset shows the magnified view of $\rho(T)$ below 40 K for Lu$_{0.925}$Yb$_{0.075}$Rh$_2$Zn$_{20}$, showing a local minimum at about 20 K. \label{LYRZ_rho_Tdep}}
\end{figure}

As shown in Fig. \ref{LYRZ_rho_Tdep}, the temperature variation of electrical resistivity for $x = 0$ (black line) and 0.075 (blue line) shows metallic behavior.
The appearance of the local minimum when substituting Yb impurities into \ce{LuRh2Zn20} indicates $c$-$f$ scattering due to the Kondo effect at Yb impurity sites. The $4f$ contribution of the resistivity, shown in Fig. 2(d) of the manuscript, was extracted as $\rho_{4f} = \rho(x = 0.075) - \rho(x = 0)$.

\section{Inelastic neutron scattering}

\begin{figure}[t!]
	\includegraphics[keepaspectratio, width=0.4\textwidth, clip]{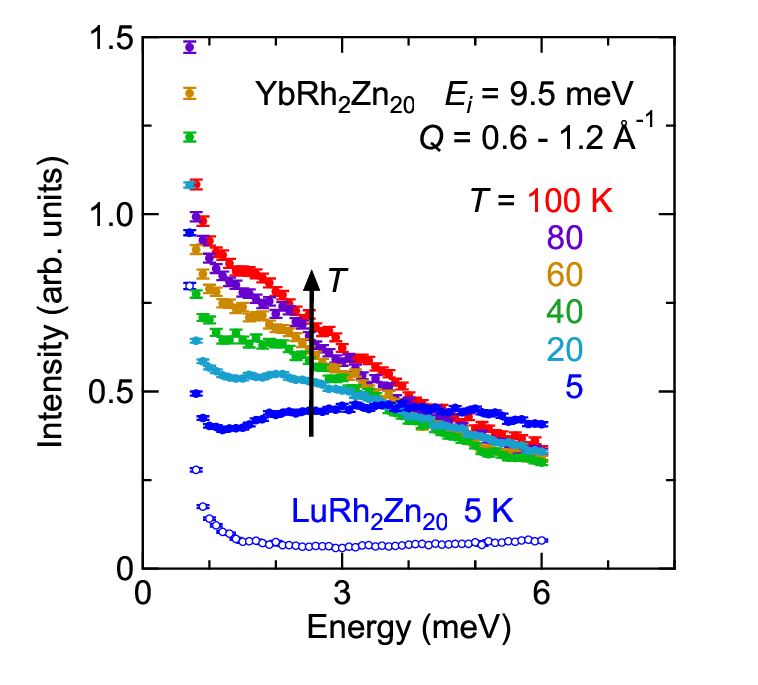}
	\caption{Raw inelastic neutron scattering  spectra of \ce{YbRh2Zn20} and \ce{LuRh2Zn20} integrated over $Q = 0.6$ - $1.2 \ \text{\AA}^{-1}$ with incident energy $E_i = 9.5 \ \mathrm{meV}$ and Fermi chopper frequency $f = 200 \ \mathrm{Hz}$. \label{YRZ_INS_rawdata}}
\end{figure}

Figure \ref{YRZ_INS_rawdata} shows the INS spectra of \ce{YbRh2Zn20} and \ce{LuRh2Zn20}. The spectrum at 5 K for the Yb system reveals a broad peak at around $\hbar\omega = 4 \ \mathrm{meV}$, shifting to $\hbar\omega = 0 \ \mathrm{meV}$ with increasing temperature. In contrast, the nonmagnetic reference compound \ce{LuRh2Zn20} shows no significant inelastic scattering.
Thus, the broad peak in \ce{YbRh2Zn20} results from the magnetic scattering of Yb ions. As described in the main manuscript, the temperature evolution of the INS spectra can be interpreted as paramagnetic quasielastic scattering due to the Kondo effect.


%



\nocite{*}
\addtocontents{toc}{\protect\setcounter{tocdepth}{0}}
\bibliography{bibliography_SM}

%
%
%
%
%

